\documentclass[twocolumn,superscriptaddress,prx]{revtex4-2}

\usepackage[utf8]{inputenc}
\usepackage[english]{babel}
\usepackage[T1]{fontenc}
\usepackage{amsmath}
\usepackage{amssymb}
\usepackage{hyperref}

\usepackage{siunitx}
\usepackage{physics}
\usepackage{cancel}

\usepackage[export]{adjustbox}
\usepackage{graphicx}
\usepackage{xcolor}
\usepackage{ulem}

\definecolor{mypurple}{rgb}{0.49,0.18,0.56}
\definecolor{mygreen}{rgb}{0,0.5,0}
\definecolor{mygold}{rgb}{0.93,0.59,0.13}
\definecolor{myblue}{rgb}{0,0,0.75}
\definecolor{mymagenta}{cmyk}{0,1,0,0.12}
\definecolor{lavender}{rgb}{0.57,0.39,0.80}

\newcommand{\fdrv}{\ensuremath{f_\text{drv}}}
\newcommand{\mass}{\ensuremath{m}}
\newcommand{\coupling}{\ensuremath{J}}
\newcommand{\frot}{\ensuremath{f_\text{rot}}}

\begin{document}
\title{Engineering a U(1) lattice gauge theory in classical electric circuits}
\author{Hannes Riechert}
\affiliation{Universit\"at Heidelberg, Kirchhoff-Institut f\"ur Physik, Im Neuenheimer Feld 227, 69120 Heidelberg, Germany}
\author{Jad C.~Halimeh}
\affiliation{INO-CNR BEC Center and Department of Physics, University of Trento, Via Sommarive 14, I-38123 Trento, Italy}
\author{Valentin Kasper}
\affiliation{ICFO - Institut de Ciencies Fotoniques, The Barcelona Institute of Science and Technology,
Av.~Carl Friedrich Gauss 3, 08860 Castelldefels (Barcelona), Spain}
\affiliation{Department of Physics, Harvard University, Cambridge, MA, 02138, USA}
\author{Landry Bretheau}
\affiliation{Laboratoire de Physique de la Mati\`ere condens\'ee, CNRS, Ecole Polytechnique, Institut Polytechnique de Paris, 91120 Palaiseau, France}
\author{Erez Zohar}
\affiliation{Racah Institute of Physics, The Hebrew University of Jerusalem, Givat Ram, Jerusalem 91904, Israel}
\author{Philipp Hauke}
\affiliation{INO-CNR BEC Center and Department of Physics, University of Trento, Via Sommarive 14, I-38123 Trento, Italy}
\author{Fred Jendrzejewski}
\affiliation{Universit\"at Heidelberg, Kirchhoff-Institut f\"ur Physik, Im Neuenheimer Feld 227, 69120 Heidelberg, Germany}
\date{\today}
\begin{abstract}
Lattice gauge theories are fundamental to such distinct fields as particle physics, condensed matter, and quantum information science. Their local symmetries enforce the charge conservation observed in the laws of physics. Impressive experimental progress has demonstrated that they can be engineered in table-top experiments using synthetic quantum systems. However, the challenges posed by the scalability of such lattice gauge simulators are pressing, thereby making the exploration of different experimental setups desirable. Here, we realize a U(1) lattice gauge theory with five matter sites and four gauge links in classical electric circuits employing nonlinear elements connecting LC oscillators. This allows for probing previously inaccessible spectral and transport properties in a multi-site system. We directly observe Gauss's law, known from electrodynamics, and the emergence of long-range interactions between massive particles in full agreement with theoretical predictions. Our work paves the way for investigations of increasingly complex gauge theories on table-top classical setups, and demonstrates the precise control of nonlinear effects within metamaterial devices.
\end{abstract}
\maketitle

Local symmetries provide a mathematical framework to describe emergent behavior from a small set of microscopic rules. Paradigmatic examples are topological phases of matter, which can emerge as ground states of an extensive set of commuting local operators \cite{Kitaev2003,Sachdev2019}.  In the Standard Model of particle physics all interactions between elementary particles are mediated by gauge bosons~\cite{weinberg1995quantum}. Recently, there has been a flurry of proposals and experimental implementations of lattice gauge theories in quantum many-body platforms \cite{Wiese2013, Zohar2016, dalmonte2016lattice, Martinez2016, Klco2018, Lu2019, Kokail2019, Mil2020, Schweizer2019, Gorg2018, Yang2020}. Yet, gauge invariance is as fundamental in classical physics and applications thereof. A famous example is classical electrodynamics, where gauge invariance appears as Gauss's law. Its presence in Maxwell's equations has been a guiding principle for transformation optics~\cite{Leonhardt1777,Pendry1780} based on a variational approach~\cite{GarciaMeca2013}, which has led to the experimental realization of intriguing devices such as metamaterials with negative indices of refraction~\cite{Shelby77} and invisibility cloaks~\cite{Schurig977}. In this light it is natural for gauge invariance to be investigated using classical setups that are usually less expensive and simpler to implement compared to their counterparts in quantum synthetic matter.

\begin{figure}[!t]
	\includegraphics[width=\columnwidth]{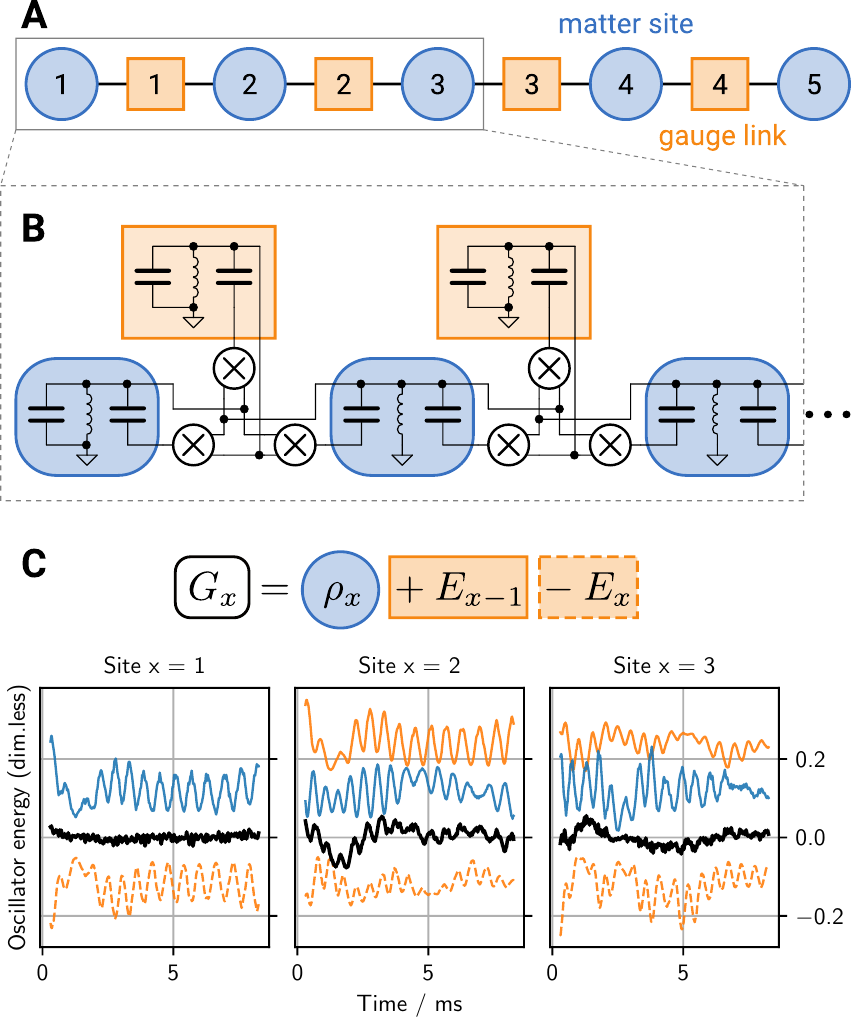}
	\caption{\textbf{Engineering a classical gauge theory.}\\
		(\textbf{\sffamily A})~Structure of a lattice gauge theory. Matter fields reside on sites, gauge fields on the links in-between.
		(\textbf{\sffamily B})~Circuit implementation with LC resonators for each site and link.  The interaction is realized by three-wave mixers as described in the main text.
		(\textbf{\sffamily C})~Measurement of Gauss's law. The Gauss law $G_x$ can be measured through the oscillator energy on each matter site (blue) and its neighboring links (orange).  Links that appear with negative sign in Gauss's law are shown inverted along y-axis (dashed). The smaller oscillation of the resulting observable (black) confirm that the local conservation laws are fulfilled. Curves are offset for clarity (see Appendix~\ref{app:conservation}).}
	\label{fig:sketch}
\end{figure}

Here, we engineer a complex metamaterial with a local U(1) gauge symmetry---the simplest continuous gauge symmetry and the basis of quantum electrodynamics---which embodies the invariance of the equations of motion under a local phase transformation. We base our setup on classical electric circuits, which have proven to be a powerful platform for studying  topological lattice structures and multidimensional metamaterials \cite{Ningyuan2015,Imhof2018,Lee2018}, as well as noise-assisted energy transport \cite{Leon2013,Leon2015}.

In the material, electric fields propagate through a chain of LC oscillators, and the U(1) symmetric coupling is engineered from three-wave mixers. The radio-frequency circuit is then described by nonlinear differential equations, similar to the ones known from  nonlinear optics. Based on this approach, we experimentally demonstrate the engineering of a lattice with nine LC oscillators that represent five matter fields and four gauge links, see Fig.~\ref{fig:sketch}. We demonstrate the high tunability of the setup, and confirm its faithful representation of the desired model using analytical models and numerical benchmark calculations. The ease of use and low cost of such classical metamaterials compared to quantum experiments open new ways to employ gauge symmetries for a wide range of materials from acoustics \cite{He2016Acoustic} over photonics \cite{Lu2014photonics} to mechanical pendulums \cite{Susstrunk2015}.

Our material can be described within the language of a classical lattice gauge theory as sketched in Fig.~\ref{fig:sketch}A. In this framework, matter fields reside on discrete sites $x$ of a 1-dimensional lattice and are coupled through links, which host the gauge fields. Within our modeling, we implement both the matter and the gauge fields by harmonic oscillators described by the complex numbers $a_x$ and $b_x$. In the appropriate rotating frame, the matter sites $a_x$ have staggered frequencies $(-1)^x\,\mass$ (see Appendix~\ref{app:hamiltonian}). Two consecutive matter sites are coupled by a gauge field $b_x$ on the link connecting them.  The coupling term is a three-wave mixing term with the interaction frequency $\coupling$. This system is described by the classical Hamiltonian
\begin{align}\nonumber
    H ={}&\coupling \Bigg\{
        \sum_\text{$x$ odd} \big( a_{x}^{\ast}\, b_{x}^{\ast}\, a_{x+1} + \text{c.\,c.} \big) \\\nonumber
    &
        + \sum_\text{$x$ even} \big( a_{x}^{\ast}\, b_{x}\, a_{x+1} + \text{c.\,c.}\big)
        \Bigg\} \\\label{eq:hamiltonian}
    &-\mass \sum_{x=1}^{l}(-1)^{x}\, \, a_{x}^{\ast}\, a_{x},
\end{align}
with equations of motion
\begin{align}\label{eq:h-eom}
  -i \dot a_x &= \pdv{H}{a_x^*}\,, & -i \dot b_x &= \pdv{H}{b_x^*}\,.
\end{align}

We realize the Hamiltonian~\eqref{eq:hamiltonian} through an array of LC oscillators, whose charge $Q_x$ (flux $\Phi_x$) represents the real (imaginary) part of the complex fields $a_x$.  Dividing out a typical energy and time scale allows us to write the Hamiltonian in units of frequency and in terms of dimensionless fields $a_x$ and $b_x$.

The Hamiltonian is invariant under the local U(1) transformation $a_x \rightarrow a_x\,e^{i\theta_x}$ where the gauge field absorbs the difference $b_x \rightarrow b_x\,e^{i (-1)^x (\theta_x-\theta_{x+1})}.$
With the gauge symmetry comes a local conserved quantity $G_x$, which is the generator of Gauss's law. Writing $\rho_x = a_x^*\,a_x$ and $E_x = -(-1)^x b_x^*\,b_x$ yields the expression $G_x = \rho_x + E_{x-1} - E_x$. Interpreting $\rho_x$ and $E_x$ as the local oscillators' energies, Gauss's law describes the conservation of the total energy on a matter site and its neighboring sites and gauge fields. Gauss's law is implemented through the capacitive coupling with a ring of three appropriately connected voltage multipliers (see Appendix \ref{app:oscillators}). The frequencies of the LC oscillators are designed such that the sum of the odd matter sites and the links is approximately equal to the frequency of the even matter sites. The small frequency detuning $\mass$ can then be isolated from the fast timescales in the rotating frame.  By averaging over fast timescales on the order of the free resonance frequency $\omega_x$, gauge-violating terms in the coupling are removed \cite{Halimeh2020,Yang2020,halimeh2020gaugesymmetry}, which allows the U(1)~gauge invariant Hamiltonian of Eq.~\eqref{eq:hamiltonian} to emerge.

The LC oscillators realizing the matter sites have a free resonance frequency of \SI{31.0(5)}{\kilo \Hz} (\SI{86(1)}{\kilo \Hz}) for odd (even) sites (see Appendix~\ref{app:oscillators}).  The links have a free resonance frequency of $\SI{55}{\kilo \Hz}$ with the possibility to be tuned. This setup results effectively in a controllable detuning of $\SI{0}{\kilo \Hz} \leq \mass \leq \SI{4}{\kilo \Hz}$ with a precision of $\SI{200}{\Hz}$. The coupling strength, depending on free resonance frequencies, is $J=\SI{0.92(5)}{kHz}$ at $m=\SI{2.5}{kHz}$. With a typical quality factor of 50 the dissipation in the resonators takes effect before the U(1) interaction. As a remedy, a positive feedback current controlled by a pickup coil in the resonator inductors is added, resembling regenerative receivers of early radio technology. We observe a non-trivial energy exchange between matter sites as shown in Fig.~\ref{fig:sketch}C. The local symmetry enforces concerted dynamics of matter sites and their neighboring links, such that the measured Gauss law has only small variations. This observation quantifies the weak violation of local gauge invariance in our metamaterial.

\begin{figure}
    \centering
	\includegraphics[width=\columnwidth]{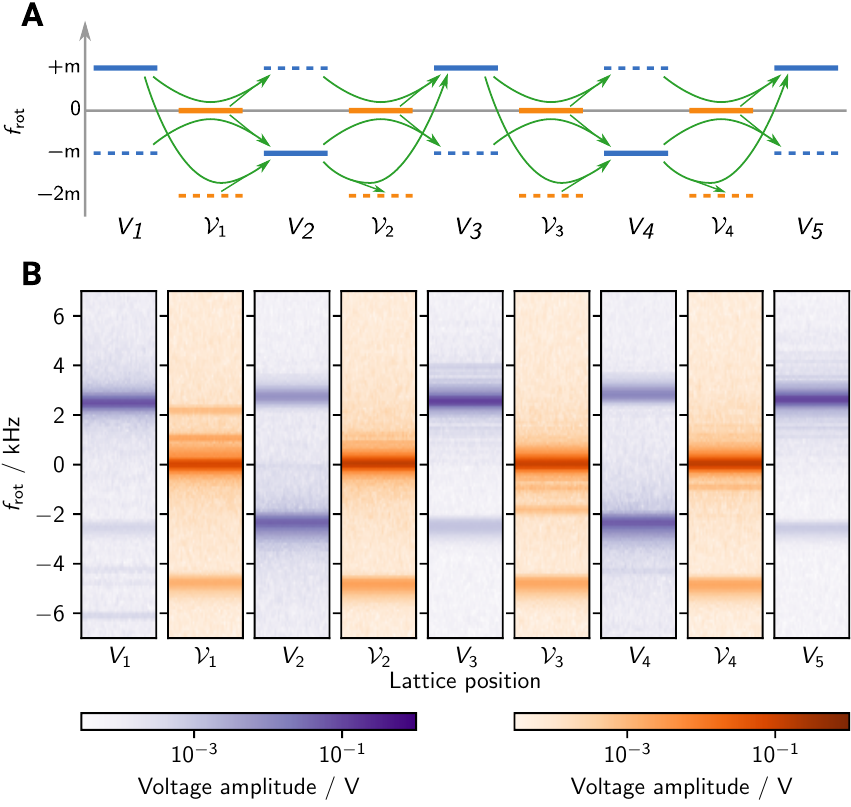}
    \caption{\textbf{Site-resolved spectrum.}
    (\textbf{\sffamily A}) Level scheme of the Hamiltonian in the regime of large mass $\mass = \SI{2.5}{kHz}$ compared to the coupling strength $\coupling=\SI{0.92}{kHz}$. The bare frequencies of the matter sites (blue) have a resonance at $\frot = -(-1)^x m$ and the links (orange) at $\frot = 0$, shown as straight lines. They are dressed by the gauge invariant coupling, which allows excitations to move between matter sites (green arrows). The coupling results in weak spectral lines at $\frot = (-1)^x\,m$ for the matter field, which are mirrored on the links through weak spectral lines at $\frot = - 2m$ due to the Gauss law.
    (\textbf{\sffamily B}) Observed frequency spectrum of oscillator voltages. The positions of spectral lines compare well to the predicted level structure from perturbation theory.}
    \label{fig:level-structure}
\end{figure}

To analyze the non-trivial dynamics, we investigate the spectral properties of the chain. We set the initial conditions to $Q_x(t=0) = 0$ and the initial flux is chosen such that the oscillators start with an amplitude of \SI{\sim 0.7}{V}. After initialization, \SI{8.3}{ms} long time traces of the voltage signals of all resonators are available for spectral analysis as shown in Fig.~\ref{fig:level-structure}. The observed spectra can be well understood perturbatively as they are obtained in the regime of large detuning. The spectra of  the matter sites contain two frequency components of different strengths. The stronger frequency originates from the free resonance of the LC oscillators. The second and weaker one originates from the pertubative interactions with the gauge links. Gauss's law implies that the gauge field has to match the appearance of a second frequency component in the matter field. This appearance is clearly visible in our spectra and is observed at the predicted frequency.

\begin{figure}
    \centering
    \includegraphics[width=\columnwidth]{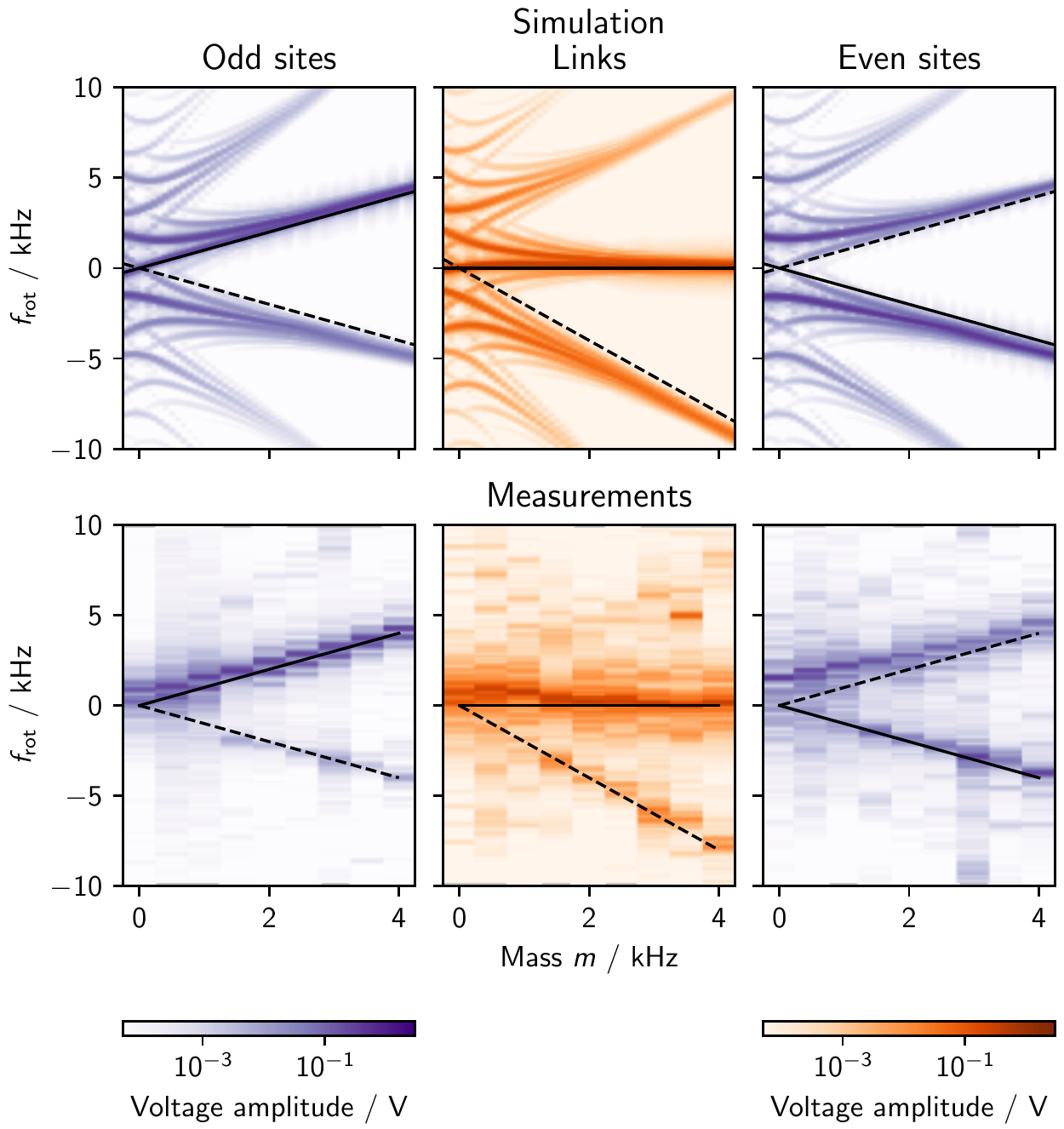}
    \caption{\textbf{Mass dependence of the lattice gauge theory spectrum.} A spectral analysis of our model in Eq.~\eqref{eq:hamiltonian} with five matter sites, as a function of the matter field mass $\mass$, is presented in terms of experimental measurements and numerical simulations. All spectral features observed in the experiment are well reproduced by the numerics. Agreement with first-order perturbation theory in the thermodynamic limit (black lines) confirms that our implementation is sufficiently large to render finite-size effects insignificant.}
    \label{fig:spetrum-detuning}
\end{figure}

We employed the circuit to systematically investigate the dependence of the spectrum on the mass, as shown in Fig.~\ref{fig:spetrum-detuning}. For large mass, we observe experimentally a linear dependence of the spectral lines, which is well explained by first-order perturbation theory in the thermodynamic limit (see Appendix \ref{app:pt-nondriven}).  This agreement indicates that even at few matter sites, our lattice gauge theory faithfully reproduces the thermodynamic limit. Similar behavior has also been seen in a quantum-link-model lattice gauge theory, where few lattice sites can capture the dynamics of local observables in the thermodynamic limit \cite{VanDamme2021}. For smaller mass, we observe deviations from the perturbative predictions derived around the large-mass limit, while non-perturbative numerical simulations of the spectra yield a quantitative agreement for the salient experimental observations over the full regime.

\begin{figure*}
    \centering
    \includegraphics[width=\textwidth]{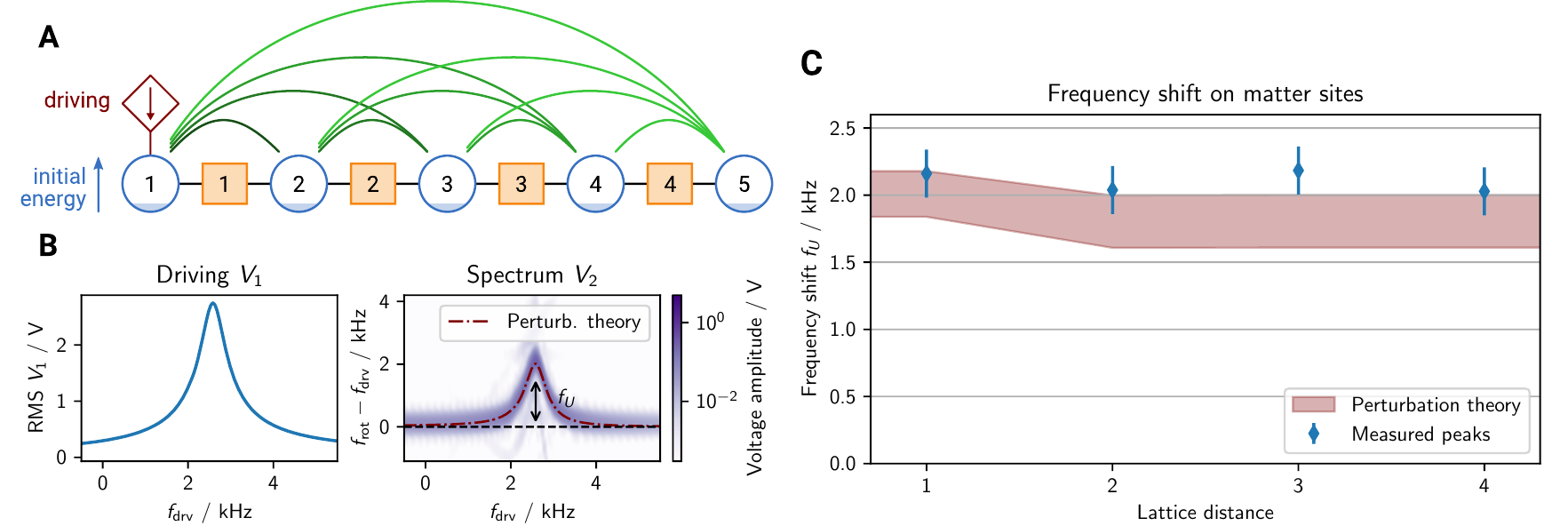}
    \caption{\textbf{Long range interactions between matter sites.}
    (\textbf{\sffamily A}) The existence of the links introduces
    The gauge invariant coupling of the matter sites leads to long-range interactions between them (green lines). To observe the resulting frequency shift, we populate the first site in a controlled fashion by driving with an alternating current of frequency $\fdrv$.
    (\textbf{\sffamily B}) The first matter site has the frequency response of a harmonic oscillator with resonance frequency $\mass = \SI{2.5(2)}{\kilo\Hz}$ and its signal is off-resonantly coupled into the second matter site with a coupling strength $\coupling=\SI{0.92(5)}{kHz}$. The response of the second matter site has a marked frequency shift, which is well explained through perturbation theory by the interaction with the first matter site (red line).
    (\textbf{\sffamily C}) We measured the frequency shifts $f_U$ as a function of lattice distance to the first site.  The observed independence of distance within the experimental uncertainties is in good agreement with perturbation theory. The uncertainties are systematically limited by the frequency resolution.}
    \label{fig:interaction}
\end{figure*}

In our metamaterial, the Gauss law implies long-range interactions between matter sites \cite{Martinez2016} as visualized in Fig.~\ref{fig:interaction}A (see Appendix \ref{app:long-range-interactions} for the derivation). They manifest as shifts of the resonance frequency. To investigate these shifts, we populate the first matter site by driving it with an alternating current. The first site has the typical frequency response of a driven harmonic oscillator with resonance at $\fdrv=+\mass$, where $\fdrv$ is the driving frequency in the rotating frame of this first site as explained in Appendix \ref{app:pt-driven}.
Amplitudes of the other matter sites are initialized to zero, while the amplitudes of dynamical links are initialized as previously. We observe a significant shift of the resonance frequency on the second matter site as shown in Fig.~\ref{fig:interaction}B.
We extended this measurement of the frequency shift $f_U$ to all matter sites as shown in Fig.~\ref{fig:interaction}C. We observe no significant decay of $f_U$ as a function of distance from the first site. The role of interactions is confirmed by good quantitative agreement between our observations and perturbation theory without free parameters. These findings show, that our experimental platform provides high control over long-range interactions through the engineering of local symmetries.

This work opens the door towards the investigation of gauge theories in electrical circuits. The realization of the Hamiltonian \eqref{eq:hamiltonian} is directly transferable to the quantum realm using superconducting circuits architectures cooled-down to 10 mK, which can be manipulated and readout with microwave signals while maintaining long enough quantum coherence. In such a platform, the required gauge invariant three-wave mixing interaction can be implemented using Josephson ring modulators based on superconducting tunnel junctions \cite{Bergeal2010}. Our work also directly offers the possibility to implement systems in higher dimensions following models that were proposed for 2D \cite{Zohar2011, Tagliacozzo2013, zohar2013quantum, Dutta2017, Zohar2017, Celi2020, Paulson2020, Ott2020} or non-Abelian systems \cite{Zohar2013-2, Banerjee2013, Tagliacozzo2012, Stannigel2013, Kasper2020Engineer, Davoudi2020, Kasper2020, Atas2021, Halimeh2021} with the exciting prospect of directly observing confinement predicted in theses theories.

\begin{acknowledgments}
    The authors are grateful for fruitful discussions with  T. Gasenzer, J. Berges, and the members of the SynQS seminar.
	This work is part of and supported by the DFG Collaborative Research Centre ``SFB 1225 (ISOQUANT)'', 
	the ERC Advanced Grant ``EntangleGen'' (Project-ID 694561),
	the ERC Starting Grant ``StrEnQTh'' (Project-ID 804305), 
	Quantum Science and Technology in Trento (Q@TN), 
	the Provincia Autonoma di Trento, 
	and
	the Excellence Initiative of the German federal government and the state governments – funding line Institutional Strategy (Zukunftskonzept): DFG project number ZUK 49/Ü. ICFO group acknowledges support from ERC AdG NOQIA, State Research Agency AEI (“Severo Ochoa” Center of Excellence CEX2019-000910-S, Plan National FIDEUA PID2019-106901GB-I00/10.13039 / 501100011033, FPI, QUANTERA MAQS PCI2019-111828-2 / 10.13039/501100011033), Fundació Privada Cellex, Fundació Mir-Puig, Generalitat de Catalunya (AGAUR Grant No. 2017 SGR 1341, CERCA program, QuantumCAT \ U16-011424, co-funded by ERDF Operational Program of Catalonia 2014-2020), EU Horizon 2020 FET-OPEN OPTOLogic (Grant No 899794), and the National Science Centre, Poland (Symfonia Grant No. 2016/20/W/ST4/00314), Marie Sk\l odowska-Curie grant STREDCH No 101029393, “La Caixa” Junior Leaders fellowships (ID100010434),  and EU Horizon 2020 under Marie Sk\l odowska-Curie grant agreement No 847648 (LCF/BQ/PI19/11690013, LCF/BQ/PI20/11760031,  LCF/BQ/PR20/11770012).
    F.\,J. acknowledges the DFG support through the Emmy-Noether grant (project-id 377616843).
	P.\,H. acknowledges the Google Research Scholar Award ProGauge. 
	E.\,Z. was  supported by the Israel Science Foundation (grant No. 523/20).
	V.\,K. received support from the ”la Caixa” Foundation (ID 100010434) and from the European Union’s Horizon 2020 research and innovation programme under the  Marie Skłodowska-Curie grant agreement No 847648 with fellowship code LCF/BQ/PI20/11760031.
\end{acknowledgments}

%

\onecolumngrid


\clearpage
\section*{Supplemental Material}
\twocolumngrid
\appendix
\section{Hamiltonian derivation}\label{app:hamiltonian}
The equations of motion for the circuit sketched in Fig.~\ref{fig:sketch}B written as Kirchhoff current laws are
\begin{subequations}\label{eq:kcl}
\begin{align}
    I_x &= \frac{1}{L_x} \Phi_x + 2 C \ddot \Phi_x
    \nonumber \\ &\hphantom{=}
    - \frac{C}{V_\text{ref}} \left( \dv{t} \dot\Phi_{x-1}^\prime \dot\Phi_{x-1}
    + \dv{t} \dot\Phi_x^\prime \dot\Phi_{x+1} \right)\,, \\
    I_x^\prime &= \frac{1}{L_x^\prime} \Phi_x^\prime + 2 C \ddot \Phi_x^\prime
    - \frac{C}{V_\text{ref}} \dv{t} \dot\Phi_x \dot\Phi_{x+1}\,.
\end{align}
\end{subequations}
On the left hand side are external currents $I_x^{(\prime)}$, where primed quantities denote the links. The right hand side is formulated in terms of magnetic fluxes $\Phi_x^{(\prime)}$ which are integrals of the voltage signal $\dot\Phi=V$. The index $x$ runs from $1$ to $l$ for sites and from $1$ to $l-1$ for links.  $V_\text{ref}=\SI{10.2(3)}{V}$ is the internal reference of the voltage multipliers. Upon setting $I_x^{(\prime)}$ to zero, which holds true when no external drive is connected, these equations arise from the following Lagrange function with generalized coordinates~$\Phi_x^{(\prime)}$:
\begin{equation}\label{eq:u1-lagrange}
  \mathcal L = \frac{1}{2} \sum_{x,x^\prime} \left( 2C \dot\Phi_x^2 - \frac{1}{L_x} \Phi_{x}^2 \right)
  - \sum_{x=1}^{L-1} \frac{C}{V_\text{ref}} \dot\Phi_x \dot\Phi_x^\prime \dot\Phi_{x+1}.
\end{equation}

Because the interaction is in the momentum terms, the conjugate momenta $Q=\partial\mathcal L / \partial\dot\Phi$, which are needed to transform to the Hamiltonian, are non-trivial to invert for $\dot\Phi(Q)$.  In the small coupling approximation, i.e., assuming typical voltages $\langle\dot\Phi\rangle$ to be much smaller than the reference voltage $V_\text{ref}$, the conjugate momentum is the local oscillator's electric charge $Q_x \approx 2C\dot\Phi_x$.

After defining the scales $V_0=\SI{1}{V}$ for voltage and $f_0=\omega_0/2\pi=\SI{60}{kHz}$ for time, we change to dimensionless flux $\bar\Phi = \omega_0\,\Phi/V_0$ and charge $\bar Q = Q/2C V_0$. The corresponding Hamiltonian can be expressed in dimensionless units after dividing by the energy scale $2C V_0^2$:
\begin{align}
  \bar H = \frac{1}{2} \sum_{x,x^\prime} \left(
  \omega_x^2\, \bar \Phi_x^2 + \bar Q_x^2 \right)
  + \sum_x \frac{V_0}{2V_\text{ref}} \bar Q_x \bar Q_x^\prime \bar Q_{x+1}\,.
\end{align}
Here, the free resonance frequencies $1/\sqrt{2 C L_x}$ of the LC oscillators are normalized to $\omega_x=1/(\omega_0\, \sqrt{2 C L_x})$.   Using dimensionless time $\tau=t\omega_0$ allows Hamilton's equations to appear unchanged.

The complex variables are defined from dimensionless quantities as
\begin{equation}\label{eq:u1-complex}
  \begin{aligned}
    a_x &= \tfrac{1}{\sqrt{2\omega_x}} (\bar Q_x + i \omega_x \bar\Phi_x), &
    b_x &= \tfrac{1}{\sqrt{2\omega_x^\prime}} (\bar Q_x^\prime + i \omega_x^\prime \bar\Phi_x^\prime)\,,
  \end{aligned}
\end{equation}
with complex conjugates as canonical momenta.  The Poisson brackets with respect to the old variables $(\Phi,Q)$ are $\{a_x,a_y^*\}=i\delta_{xy}$, requiring the additional $i$ in Hamilton's equations (Eq.~\eqref{eq:h-eom} in main text).  The Hamiltonian in the new variables is
\begin{equation}\label{eq:h-with-violation}
  \begin{aligned}
    H =& \sum_{x=1}^l \omega_x\, a_x^*\, a_x + \sum_{x=1}^{l-1} \omega_x^\prime\, b_x^*\, b_x \\
    & + \sum_{x=1}^{l-1} J_x\, (a_x+a_x^*) (b_x + b_x^*) (a_{x+1} + a_{x+1}^*)
  \end{aligned}
\end{equation}
with coupling strength
\begin{equation}\label{eq:u1-coupling}
  J_x = \frac{1}{4} \sqrt{\omega_x \omega_x^\prime \omega_{x+1} / 2\,}\, \frac{V_0}{V_\text{ref}}.
\end{equation}
Only two of the coupling terms in Eq.~\eqref{eq:h-with-violation} are invariant under the local U(1) transformation introduced in the main text and discussed in Appendix~\ref{app:conservation}).  All other terms can be separated in a rotating frame with the non-unique staggered tuning
\begin{equation}\label{eq:staggered-tuning}
    \omega_x - \omega_{x+1} = (-1)^x\, (\omega_x^\prime - \delta_x)
\end{equation}
requiring $\delta_x$ to be much smaller than the resonance frequencies.
The rotating frame is then
\begin{subequations}
\begin{align}
    a_x &\rightarrow a_x\,e^{i (\omega_x + (-1)^x m_x) \tau}\,, \\
    b_x &\rightarrow b_x\, e^{i \omega_x^\prime \tau}\,,
\end{align}
\end{subequations}
using site based detunings $m_{x+1}=\delta_x-m_x$, $m_1=\delta_1$.  The simplest allowed configuration is all $m=m_x=\delta_x$ having the same value.  In the rotating frame, and with $J_x=J$ and $m_x=m$ independent of $x$, the Hamiltonian takes on the form of Eq.~\eqref{eq:hamiltonian} in the main text.  By dividing the equations of motion by the time scale $1/\omega_0$, all measurements of the Hamiltonian are in units of frequency:
\begin{equation}
    \dv{\tau} a_x = i \pdv{H}{a_x^*}
    \quad\Leftrightarrow\quad
    \dv{t} a_x = i \omega_0 \pdv{H}{a_x^*}\,.
\end{equation}
Similarly the time scale can be absorbed into the mass $\mass$ and coupling $\coupling$ to give them and all figure axes intuitive units.

\textit{Driving.}
For the Hamiltonian description, we have set the external currents $I_x^{(\prime)}(t)$ to zero, but for the driven model, these have to be taken into account.  This is achieved by adding a term to the equations of motion \eqref{eq:h-eom}:
\begin{subequations}
\begin{align}\label{eq:eom-driven}
    \dv{a_x}{\tau} &= i\pdv{H}{a_x^*} + \hat I_x(\tau)\,, \\
    \dv{b_x}{\tau} &= i\pdv{H}{b_x^*} + \hat I_x^\prime(\tau)\,.
\end{align}
\end{subequations}
Here, $\hat I_x^{(\prime)}(\tau)$ is the dimensionless external current in the rotating frame.  For the first site, which is the only one in our experiment that is connected to an external drive, the calculation from $I_x(t)$ is as follows:
\begin{equation}
    \hat I_1(\tau) = \frac{I_1(\tau/\omega_0)}{2 \sqrt{2\omega_1}\, \omega_0 CV_0} e^{-i (\omega_1 - m_1) \tau}\,.
\end{equation}
A harmonic driving signal $I_1(t)=I \sin(2\pi f_\text{ext}t)$ takes in the rotating wave approximation the form
\begin{align}
    \hat I_1 &= \frac{I}{\sqrt{2\omega_1}\, \omega_0 2CV_0}\; \sin{\left(\tfrac{f_\text{ext}}{f_0} \tau\right)}\, e^{-i (\omega_1 - m_1) \tau} \nonumber\\
    &\approx \frac{1}{2i} \frac{I}{\sqrt{2\omega_1}\, \omega_0 2CV_0}\, e^{(f_\text{ext}/f_0-\omega_1+m_1)\tau}\,.
\end{align}
In the main text, we use $\fdrv=f_\text{ext}/f_0-\omega_1+m_1$ to denote the external driving frequency, in the rotating frame.

\textit{Dissipation.} We want to inspect features of the system with external driving close to resonance.  Without taking into account dissipation, numerical results diverge in these regimes.  As a remedy, we add an empirical dissipation term to the equations of motion, 
\begin{subequations}
\begin{align}
    \dv{a_x}{\tau} &= i\pdv{H}{a_x^*} + \hat I_x(\tau) - k a_x\,, \\
    \dv{b_x}{\tau} &= i\pdv{H}{b_x^*} + \hat I_x^\prime(\tau) - k b_x\,.
\end{align}
\end{subequations}
The dissipation term is used only for numerical results of the driven system shown in Fig.~\ref{fig:driving-full}.

\textit{Simulation.}  The equations of motion above, including driving and dissipation terms, are integrated with standard numerical solvers starting from initial conditions like in the experiment.  The parameters of mass and coupling are based on tuning parameters.  Only the dissipation strength is treated as free parameter and adjusted to $k=0.0045$ such that observations match the simulations in Fig.~\ref{fig:driving-full}.  Numerical results have negligible violation of local conservation laws.


\section{Properties of the LC oscillators and multipliers}\label{app:oscillators}
\begin{figure}
    \centering
    \includegraphics[width=\columnwidth]{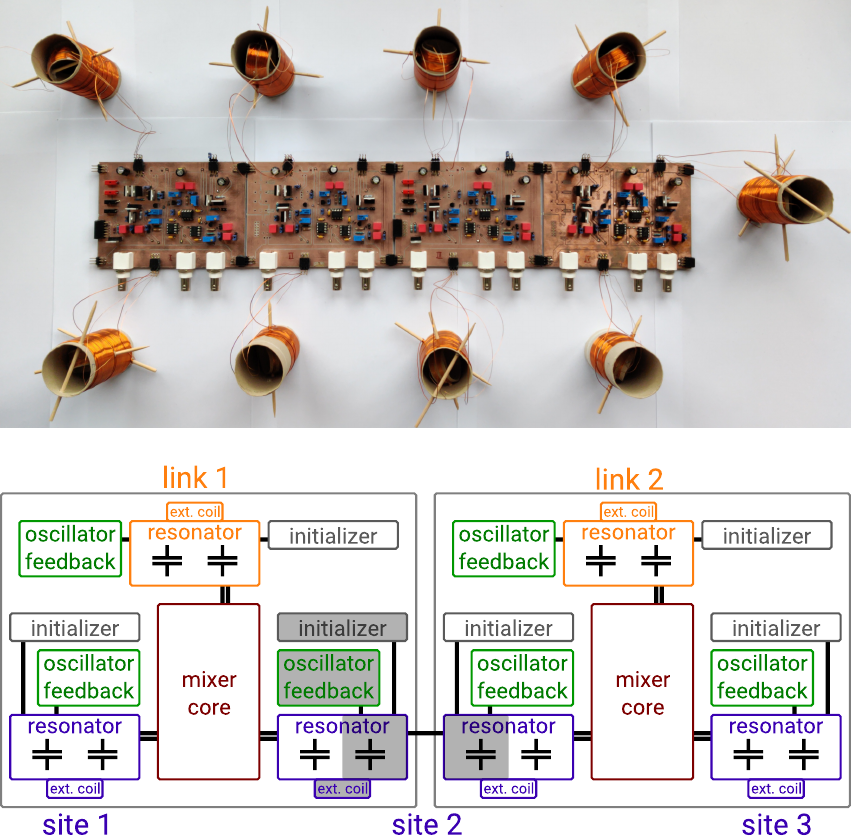}
    \caption{\textbf{Setup.} Photograph of the circuit with 5 sites.  Milled circuit boards can be chained together for arbitrarily large lattices.  Inside the handmade coils, secondary feedback and tuning coils are visible.  The illustration below indicates the general content and connections on the circuit boards.}
\end{figure}

This section describes implementation details of the circuit and lists some device properties.  The overall design frequency of the circuit between \SI{10}{kHz} and \SI{100}{kHz} has a number of advantages: it allows cheap integrated circuits to be used, does not require high-frequency aware design of the circuit, and simplifies the recording of fully sampled time traces of the dynamics.

\textit{Capacitors.} All capacitors in the circuit schematic (Fig.~\ref{fig:sketch}B) are polypropylene film capacitors and have the same value of $C=\SI{10}{nF} \pm 2\%$.  Similar to the design of some topological metamaterials~\cite{Ningyuan2015}, the coupling capacitance doubles as the on-site capacitance and the small coupling limit is achieved by a scale factor in the multipliers.

\textit{Inductors.} The inductors $L_x$ are handmade to match resonance frequencies $\omega=1\sqrt{2CL_x}$ in the range of \SI{31.0(5)}{kHz} to \SI{85(1)}{kHz}.  The main inductors are multilayer coils with a diameter of \SI{4.0(3)}{cm} and lengths between \SI{5}{cm} to \SI{9}{cm}.  The winding numbers range between \SI{80} and \SI{270}.  Inside the coils secondary coils for the feedback circuit are mounted.  The can be rotated to tune the amount of feedback.  The inductors for link resonators hold next to the feedback coil another coil, that is connected in series with the outer coil, and allows the links to be tuned from \SI{53}{kHz} to \SI{64}{kHz}.


\textit{Mixer core.} The interaction term of the U(1) Hamiltonian is a $Q_x Q_x^\prime Q_{x+1}$ coupling (see Eq.~\ref{eq:u1-lagrange}), which is symmetric under exchange and therefore the implementation must have the same symmetry under exchange of its connectors.  Furthermore the voltage of all additional nodes introduced by the interaction need to be fully determined by the site variables, in order to not introduce additional canonical variables. The Kirchhoff current laws in Eq.~\eqref{eq:kcl} state the interaction term as voltage multiplication (since $\dot\Phi=V$) that is coupled capacitively into the site with the additional time derivative.  In our circuit the voltage multiplication is achieved by explicitly inserting analog voltage multipliers (IC part number AD633), also called mixers.  The voltage multipliers in our circuit support an output voltage between $\pm\SI{10}{V}$. Multipliers have an internal scale appearing as $V_\text{ref}=\SI{10}{V}$ in the equations of motion.  Each interaction term in the KCL requires its own multiplier which sums up to three multipliers per link. The resulting coupling strength $J$ is given by Eq.~\eqref{eq:u1-coupling} and depends on $V_\text{ref}$ as well as the free resonance frequencies of the resonators. For the measurement of the interactions in Fig.~\ref{fig:interaction} with $m=\SI{2.5(2)}{kHz}$, it evaluates to $J=\SI{0.92(5)}{kHz}$.

\textit{Initializers.}  Initial conditions of the circuit are set by cutting of external currents at $t=0$.  External current are supplied by one MOSFET per oscillator that isolates the circuit from the external source at $t>0$.  Initial currents are tunable with on-board potentiometers.

\textit{Feedback.}  By adding feedback, dissipation effects can be reduced.  Typical quality factors of our oscillators are between \SI{50}{} to \SI{80}{}.  Longer time traces than this dissipation time scale are required to reach the frequency resolution of the spectra shown above.  Each oscillator is connected to a biased bipolar junction transistor that couples the signal picked up by the feedback coil back into the oscillator.  Careful tuning of the feedback strength is required to achieve both long time traces and small gauge violation.


\section{Conservation laws}\label{app:conservation}
As discussed in the main text, our Hamiltonian has the following local continuous symmetry, parameterized by the phase $\theta_x$:
\begin{subequations}
\begin{align}
    a_x &\rightarrow a_x\,e^{i\theta_x}\,, \\
    b_x &\rightarrow b_x\,e^{i (-1)^x (\theta_x-\theta_{x+1})}\,.
\end{align}
\end{subequations}
This symmetry transformation is a canonical transformation of the complex variables and as such can be expressed using Poisson brackets w.\,r.\,t. a generator function $G_x$, that depends on the complex variables local to $x$ \cite{Goldstein}, 
\begin{equation}\label{eq:g-infinitesimal}
    -i\, \partial z = \{ z,\, G_x \}\, \dd \theta_x\,,
\end{equation}
where $z$ denotes a vector of all canonical variables $a_x$ and $b_x$ (not including their complex counterpart), and $\partial z$ is their change under the infinitesimal gauge transformation.  For this purpose, the Poisson brackets are defined as
\begin{equation}
\begin{aligned}
    \{ f,\, g \} =& \hphantom{+}\, \sum_{x=1}^l \left( \pdv{f}{a_x} \pdv{g}{a_x^*} - \pdv{f}{a_x^*} \pdv{g}{a_x} \right) \\
    & + \sum_{x=1}^{l-1} \left( \pdv{f}{b_x} \pdv{g}{b_x^*} - \pdv{f}{b_x^*} \pdv{g}{b_x} \right).
\end{aligned}
\end{equation}
Upon integration of Eq.~\eqref{eq:g-infinitesimal}, one finds the generator
\begin{equation}
    G_x = a_x^*\, a_x + (-1)^x\, (b_{x-1}^*\, b_{x-1} + b_x^*\, b_x)\,.
\end{equation}
As a generator of a symmetry, $G_x$ is invariant under time evolution and fulfills $\{G,\,H\}=0$. In the main text we have identified the charge density $\rho_x=a_x^*\,a_x$ and electric field $E_x=-(-1)^x\,b_x^*\,b_x$ to write the generator in the very recognizable fashion of Gauss's law,
\begin{equation}
    G_x = \rho_x - E_x + E_{x-1}\,.
\end{equation}

Fig.~\ref{fig:sketch}C shows measurements in our circuit. The tuning for this measurement is $m_{1\dots 5} = 0.9(2)$, $0.9(2)$, $1.3(2)$, $1.0(2)$, $1.7(2)\SI{}{kHz}$. Measurements of the circuit produce voltage signals of the oscillators. Peaks in the voltage signal are used to reconstruct an envelope, which is also the amplitude in the rotating frame.  The voltage amplitudes are used to infer the oscillator energies.  An offset and an exponentially decaying background are removed from the energies, in order to center energy variations around zero. The energy variations $\mathcal E_x$ are then expressed in dimensionless terms using $a_x^*\,a_x=f_0\mathcal E_x/2CV_0^2f_x$ (links analogously).  The results in Fig.~\ref{fig:sketch}C show that the conservation is much better on odd sites than on even sites, because even sites are more sensitive to the feedback circuit. The violation on odd sites and on short time scales faster than \SI{0.5}{kHz} is consistent with the violation expected from the rotating wave approximation.


\section{Long-range interactions}
\label{app:long-range-interactions}

Starting from Hamiltonian \eqref{eq:hamiltonian}, we make the canonical variable change
\begin{equation}
    b_x = \sqrt{n_x^\prime} e^{i\phi_x}
\end{equation}
on the links only. Sites $a_x$ stay unchanged.
We then have the Hamiltonian
\begin{align}\nonumber
    H ={}&\coupling \Bigg\{
        \sum_\text{$x$ odd} \big( a_{x}^{\ast}\, \sqrt{n_x^\prime} e^{-i\phi_x}\, a_{x+1} + \text{c.\,c.} \big) \\\nonumber
    &
        + \sum_\text{$x$ even} \big( a_{x}^{\ast}\, \sqrt{n_x^\prime} e^{+i\phi_x}\, a_{x+1} + \text{c.\,c.}\big)
        \Bigg\} \\
    &-\mass \sum_{x=1}^{l}(-1)^{x}\, \, a_{x}^{\ast}\, a_{x}.
\end{align}
The phases $\phi_x$ can be eliminated by performing a canonical transformation on the matter fields \cite{Banuls2013,Martinez2016,Emonts2020}.  The transformation is controlled by the phase $\phi_x$ of the gauge fields:
\begin{align}
    a_x \rightarrow a_x \prod_{y<x} e^{-(-1)^y i\phi_y}.
\end{align}

The Gauss law $G_x=n_x+(-1)^x(n_{x-1}^\prime+n_x^\prime)$, where $n_x=a_x^*\,a_x$, allows us to rewrite the amplitudes of the gauge field using only amplitudes of the matter sites and the constant $G_x$:
\begin{equation}
    n_x^\prime = (-1)^x \sum_{y<x} (G_y - n_y).
\end{equation}
The variables $n_x^\prime$ and $\phi_x$ can thus be removed from the set of canonical variables and the gauge field does not appear anymore in the remaining Hamiltonian.  The $n_x^\prime$ depend on all previous occupation numbers and form long-range interaction terms.  When writing out the $n_x^\prime$, the new Hamiltonian looks like
\begin{align}\nonumber
    H ={}&\coupling \Bigg\{
        \sqrt{n_1 - G_1}\, a_1^*\, a_2 + \sqrt{G_1+G_2-n_1-n_2}\, a_2^*\, a_3 \\\nonumber
    & + \sqrt{n_1+n_2+n_3-G_1-G_2-G_3}\, a_3^*\, a_4 \\\nonumber
    & + \dots + \text{c.\,c.} \Bigg\} \\
    &-\mass \sum_{x=1}^{l}(-1)^{x}\, \, a_{x}^{\ast}\, a_{x}.
\end{align}
The coupling of one matter site to the next depends on the gauge field between them, and the gauge field is by Gauss's law the sum of all charges on the left of the chain plus some constant for the boundary (see Fig.~\ref{fig:interaction}A).  The coupling from odd to even sites increases with larger gauge fields, while the coupling from odd to even sites decreases with stronger gauge fields.  There is now a perceived directionality to the Hamiltonian, which arises from the choice in summing over Gauss's law.

Unlike conventional lattice gauge theories, e.g., the Schwinger model, our Hamiltonian does not have an electric energy term ($b_x^*\,b_x$) through which long-range interactions can arise.  In our case, the long-range interaction appears because link amplitudes $n_x^\prime$ are not constant, which in conventional lattice gauge theories is typically not the case.


\begin{widetext}
\section{Perturbation theory for non-driven system}\label{app:pt-nondriven}
We now carry out perturbation theory to provide an analytic footing for our experimental results. We start with the Hamiltonian~\eqref{eq:hamiltonian}, and assume we are in the thermodynamic limit. This reduces our model to a unit cell of two matter sites and two gauge links with periodic boundary conditions described by the Hamiltonian
\begin{align}
    H = m\big(a_1^\ast a_1-a_2^\ast a_2\big)+J\big[a_{1}^{\ast} \big(b_{1}^{\ast}+b_{2}^{\ast}\big) a_{2}+a_1\big(b_1+b_2\big)a_{2}^\ast\big].
\end{align}

We see that the dynamics of the $b$ field is the same whether on an odd or even link. As such, we can rewrite our Hamilton's equations simply as
\begin{subequations}
\begin{align}\label{eq:abODE}
    i\dd_\tau a_1 &= -ma_1-2J b^{\ast}a_{2},\\
    i\dd_\tau a_2 &= ma_2-2J b a_1,\\
    i\dd_\tau b &= -J a_1^\ast a_2.
\end{align}
\end{subequations}
The full solution of the $b$ field in terms of the $a$ fields is
\begin{align}
    b(\tau) = \mathcal{B} + iJ\int_0^\tau \dd s\,a_1^\ast(s)a_2(s).
\end{align}

Employing perturbation theory with $J/m$ as small parameter, we now solve up to third order for the $a$ and $b$ fields. The zeroth-order contribution in $J$ is
\begin{subequations}
\begin{align}
    b^{(0)}(\tau) &=\mathcal{B}\\
    a_1^{(0)}(\tau) &= \mathcal{A}_1e^{im\tau},\\
    a_2^{(0)}(\tau) &= \mathcal{A}_2e^{-im\tau}.
\end{align}
\end{subequations}
We can now solve for the first-order contribution to the $b$ field as follows:
\begin{align}
    b^{(1)}(\tau) = iJ\int_0^\tau \dd s\, a_1^{(0)\ast}(s) a_2^{(0)}(s) = iJ\mathcal{A}_1^\ast \mathcal{A}_2 \int_0^\tau \dd s\, e^{-2ims} = \frac{J}{2m}\mathcal{A}_1^\ast \mathcal{A}_2\, \big(1-e^{-2im\tau}\big).
\end{align}
The first-order contributions to the $a$ fields can be found by solving
\begin{subequations}
\begin{align}
    i\dd_\tau a_1^{(1)} &= -ma_1^{(1)} - 2J b^{(0)\ast} a_2^{(0)} = -ma_1^{(1)} -2J \mathcal{A}_2 \mathcal{B}^\ast e^{-im\tau},\\
    i\dd_\tau a_2^{(1)} &= ma_2^{(1)} - 2J b^{(0)} a_1^{(0)} = ma_2^{(1)} -2J \mathcal{A}_1 \mathcal{B} e^{im\tau},
\end{align}
\end{subequations}
the solutions to which are
\begin{subequations}
\begin{align}
    a_1^{(1)}(\tau) &= \frac{J}{m} \mathcal{A}_2 \mathcal{B}^\ast \big(e^{im\tau}-e^{-im\tau}\big),\\
    a_2^{(1)}(\tau) &= \frac{J}{m} \mathcal{A}_1 \mathcal{B} \big(e^{im\tau}-e^{-im\tau}\big).
\end{align}
\end{subequations}
Consequently, the second-order contribution to the $b$ field is
\begin{align}\nonumber
    b^{(2)}(\tau) &= iJ\int_0^\tau \dd s\, \left[ a_1^{(0)\ast} a_2^{(1)} + a_1^{(1)\ast} a_2^{(0)} \right] \\\nonumber
    &= i\frac{J^2}{m} \int_0^\tau \dd s\, \mathcal{B} \left( \lvert\mathcal{A}_1\rvert^2 - \lvert\mathcal{A}_2\rvert^2 \right) \big(1-e^{-2ims}\big) \\
    &= \frac{J^2}{m} \mathcal{B} \left( \lvert\mathcal{A}_1\rvert^2 - \lvert\mathcal{A}_2\rvert^2 \right) \left[ i\tau - \frac{1}{2m} \left(1-e^{-2im\tau} \right)\right],
\end{align}
which again carries the same frequency as the first-order contribution. We now continue to the second-order contribution in the $a$ fields by solving
\begin{subequations}
\begin{align}
    i\dd_\tau a_1^{(2)} &= -ma_1^{(2)}-2J \left[b^{(0)\ast}a_2^{(1)}+b^{(1)\ast}a_2^{(0)}\right] = -ma_1^{(2)} + \frac{J^2}{m}\mathcal{A}_1 \left(\lvert\mathcal{A}_2\rvert^2-2\lvert\mathcal{B}\rvert^2\right) \left(e^{im\tau}-e^{-im\tau}\right),\\
    i\dd_\tau a_2^{(2)} &= ma_1^{(2)}-2J \left[b^{(0)}a_1^{(1)}+b^{(1)}a_1^{(0)}\right] = ma_2^{(2)} - \frac{J^2}{m} \mathcal{A}_2 \left(\lvert\mathcal{A}_1\rvert^2+2\lvert\mathcal{B}\rvert^2\right) \left(e^{im\tau}-e^{-im\tau}\right),
\end{align}
\end{subequations}
the solutions for which are
\begin{subequations}
\begin{align}
    a_1^{(2)}(\tau) &= -\frac{J^2}{2m^2}\mathcal{A}_1 \left(\lvert\mathcal{A}_2\rvert^2-2\lvert\mathcal{B}\rvert^2\right) \left[(2im\tau-1)e^{im\tau}+e^{-im\tau}\right],\\
    a_2^{(2)}(\tau) &= \frac{J^2}{2m^2}\mathcal{A}_2 \left(\lvert\mathcal{A}_1\rvert^2+2\lvert\mathcal{B}\rvert^2\right) \left[e^{im\tau}-(2im\tau+1)e^{-im\tau}\right].
\end{align}
\end{subequations}
The third-order contribution to the $b$ field is
\begin{align}
    b^{(3)}(\tau) = iJ\int_0^\tau \dd s\, \big[a_1^{(0)\ast}a_2^{(2)}+a_1^{(1)\ast}a_2^{(1)}+a_1^{(2)\ast}a_2^{(0)}\big].
\end{align}
The third-order contribution to the $a$ fields is obtained by solving
\begin{subequations}
\begin{align}\nonumber
    i\dd_\tau a_1^{(3)} &= -ma_1^{(3)}-2J \big[b^{(0)\ast}a_2^{(2)}+b^{(1)\ast}a_2^{(1)}+b^{(2)\ast}a_2^{(0)}\big]\\
    &=-ma_1^{(3)}+c_1e^{3im\tau}+c_2e^{im\tau}+(c_3t+c_4)e^{-im\tau},\\\nonumber
    i\dd_\tau a_2^{(3)} &= ma_1^{(3)}-2J \big[b^{(0)}a_1^{(2)}+b^{(1)}a_1^{(1)}+b^{(2)}a_1^{(0)}\big]\\
    &=ma_1^{(3)}+(d_1t+d_2)e^{im\tau}+d_3e^{-im\tau}+d_4e^{-3im\tau},\\
    c_1&=\frac{J^3}{m^2}\mathcal{A}_1^2\mathcal{A}_2^\ast\mathcal{B},\\
    c_2&=-\frac{J^3}{m^2}\big[\mathcal{A}_2\mathcal{B}^\ast\big(2\lvert\mathcal{A}_1\rvert^2-\lvert\mathcal{A}_2\rvert^2+2\lvert\mathcal{B}\rvert^2\big)+2\mathcal{A}_1^2\mathcal{A}_2^\ast\mathcal{B}\big],\\
    c_3&=2i\frac{J^3}{m}\mathcal{A}_2\mathcal{B}^\ast\big(2\lvert\mathcal{A}_1\rvert^2-\lvert\mathcal{A}_2\rvert^2+2\lvert\mathcal{B}\rvert^2\big),\\
    c_4&=\frac{J^3}{m^2}\big[\mathcal{A}_2\mathcal{B}^\ast\big(2\lvert\mathcal{A}_1\rvert^2-\lvert\mathcal{A}_2\rvert^2+2\lvert\mathcal{B}\rvert^2\big)+\mathcal{A}_1^2\mathcal{A}_2^\ast\mathcal{B}\big],\\
    d_1&=-2i\frac{J^3}{m}\mathcal{A}_1\mathcal{B}\big(\lvert\mathcal{A}_1\rvert^2-2\lvert\mathcal{A}_2\rvert^2+2\lvert\mathcal{B}\rvert^2\big),\\
    d_2&=\frac{J^3}{m^2}\big[\mathcal{A}_1\mathcal{B}\big(\lvert\mathcal{A}_1\rvert^2-2\lvert\mathcal{A}_2\rvert^2+2\lvert\mathcal{B}\rvert^2\big)-\mathcal{A}_1^\ast\mathcal{A}_2^2\mathcal{B}^\ast\big],\\
    d_3&=-\frac{J^3}{m^2}\big[\mathcal{A}_1\mathcal{B}\big(\lvert\mathcal{A}_1\rvert^2-2\lvert\mathcal{A}_2\rvert^2+2\lvert\mathcal{B}\rvert^2\big)-2\mathcal{A}_1^\ast\mathcal{A}_2^2\mathcal{B}^\ast\big],\\
    d_4&=-\frac{J^3}{m^2}\mathcal{A}_1^\ast\mathcal{A}_2^2\mathcal{B}^\ast,
\end{align}
\end{subequations}
the solutions for which are
\begin{subequations}
\begin{align}
    a_1^{(3)}(\tau) &= -\frac{c_1}{2m}e^{3im\tau}-\left[ic_2t+\frac{J^3}{m^3}\mathcal{A}_2\mathcal{B}^\ast\big(2\lvert\mathcal{A}_1\rvert^2-\lvert\mathcal{A}_2\rvert^2+2\lvert\mathcal{B}\rvert^2\big)\right]e^{im\tau}+\frac{2mc_3\tau+2mc_4-ic_3}{4m^2}e^{-im\tau},\\
    a_2^{(3)}(\tau) &= -\frac{2md_1\tau+2md_2+id_1}{4m^2}e^{im\tau}-\left[id_3\tau-\frac{J^3}{m^3}\mathcal{A}_1\mathcal{B}\big(\lvert\mathcal{A}_1\rvert^2-2\lvert\mathcal{A}_2\rvert^2+2\lvert\mathcal{B}\rvert^2\big)\right]e^{-im\tau}+\frac{d_4}{2m}e^{-3im\tau}.
\end{align}
\end{subequations}

\begin{figure*}
    \centering
    \includegraphics[width=\textwidth]{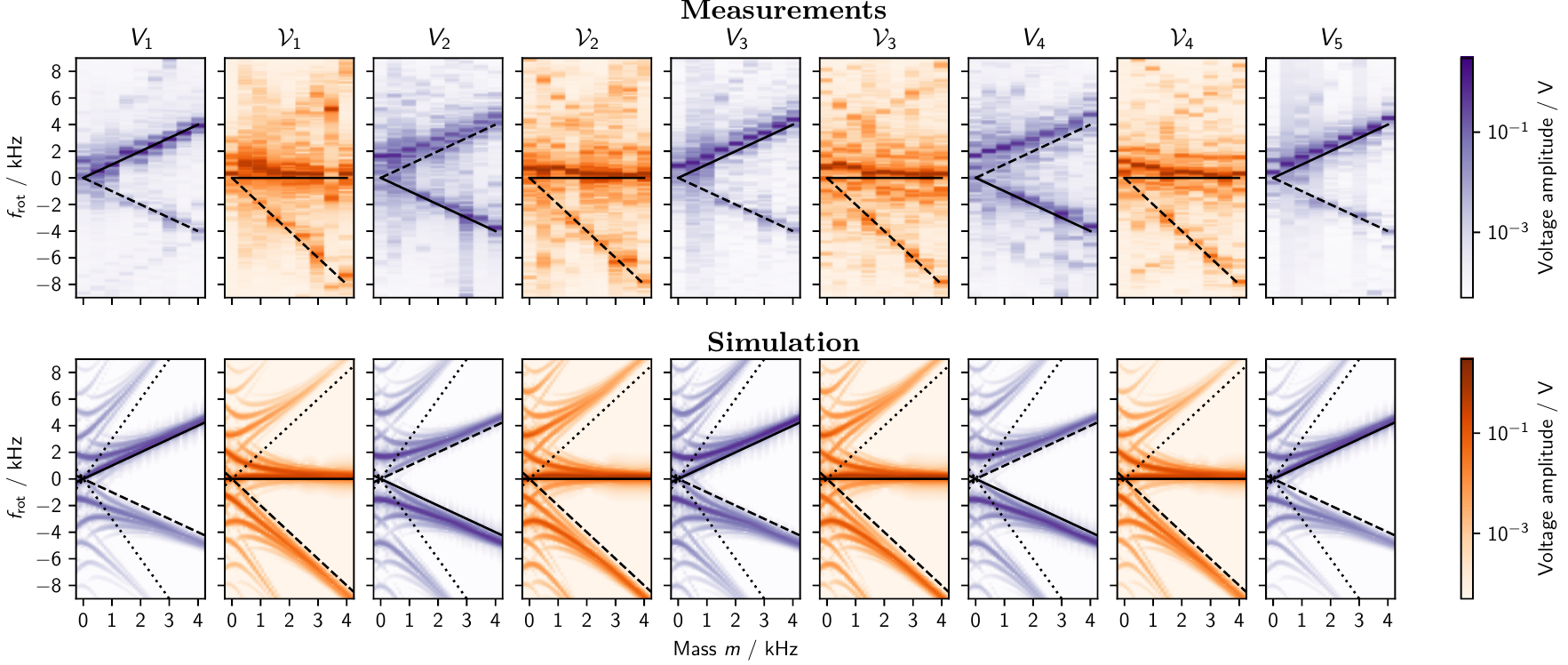}
    \caption{Spectrum of the 5-site lattice depending on detuning $\Delta=\Delta_x$.  Dashed lines indicate frequency components of first order perturbation theory.  Simulation of the U(1) Hamiltonian shows higher order components which are also discernible in the measurements.  Dotted lines in sites indicate $\pm 3m$ frequency components and in links $+2m$ components from higher orders of perturbation theory.}
    \label{fig:spetrum-detuning-full}
\end{figure*}

The results of the perturbation theory are compared to the experimental data and the numerical predictions of the spectrum in Fig.~\ref{fig:spetrum-detuning-full}.
\end{widetext}


\section{Perturbation theory for driven system}\label{app:pt-driven}
\begin{figure}
    \centering
    \includegraphics[width=\columnwidth]{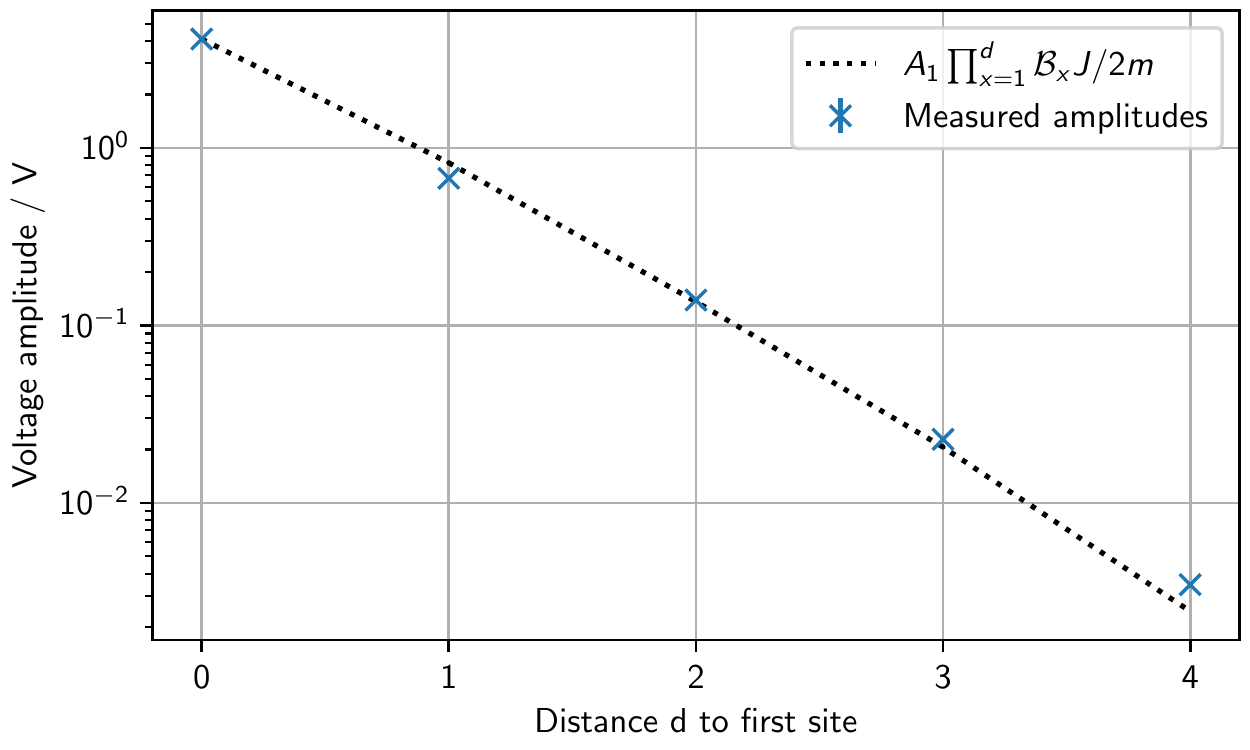}
    \caption{Observed exponential decline of oscillator voltage amplitudes along lattice when driving first site and comparison to the next neighbor coupling mediated by links calculated in Appendix~\ref{app:pt-nondriven}. This exponential decline is used as hierarchy for the perturbation theory in the driven case.}
    \label{fig:interaction-hierarchy}
\end{figure}

For the perturbation theory in the driven case, we assume a setup as described in the main text: Only the links are initialized at $t=0$ and sites are left empty.  The first site is driven with a harmonic signal, see Eq.~\eqref{eq:eom-driven}.  We know from measurements that the driving signal propagates onto the other sites with exponentially decreasing amplitude over distance (Fig.~\ref{fig:interaction-hierarchy}). The reason for this are the staggered site frequencies, such that the driving is off-resonant at least at every second site.  This motivates the approximation that the back-action of the second onto the first site and the back-action of the third onto the second site can be neglected.  Then the equations of motion of the second site become linear.  With the exponential structure of the amplitudes, this procedure can be repeated along the lattice.

Without back-action of the second site, the first site relaxes to match the driving signal $a_1 = A e^{i\nu \tau}$, $\nu=\fdrv/f_0$.  The equations of motion for the link and second site $a_{x+1}$ (without terms including the third site) form a two-level system:
\begin{equation}\label{eq:dynamic-eom-n1solved}
  \begin{pmatrix} \dot b_x \\ \dot a_{x+1} \end{pmatrix}
  = i\begin{pmatrix}
  0 & J a_x^*(\tau) \\
  J a_x(\tau) & -m
  \end{pmatrix}
  \begin{pmatrix} b_x \\ a_{x+1} \end{pmatrix}.
\end{equation}
With the variable change
\begin{subequations}
\begin{align}
    b_x &\rightarrow b_x^\prime & b &= e^{-\frac{1}{2} i (m + \nu) \tau} b^\prime \\
    a_{x+1} & \rightarrow a_{x+1}^\prime &\qquad a_{x+1} &= e^{-\frac{1}{2} i (m - \nu) \tau} a_{x+1}^\prime
\end{align}
\end{subequations}
the time dependence and energy offset are removed:
\begin{equation}
  \begin{pmatrix} \dot b_x^\prime \\ \dot a_{x+1}^\prime \end{pmatrix}
  = i\begin{pmatrix}
  +\tfrac{1}{2} (\nu + m) & J A^* \\
  J A & -\tfrac{1}{2} (\nu + m)
  \end{pmatrix}
  \begin{pmatrix} b_x^\prime \\ a_{x+1}^\prime \end{pmatrix}.
\end{equation}
Eigenvalues of the matrix are
\begin{equation}
    \mu_o(\nu, A) = \pm\frac{1}{2}\sqrt{(\nu+m)^2 + 4 J^2 |A|^2}
\end{equation} and solutions have the shape
\begin{subequations}
\begin{align}
    b_x &= c_1 e^{+i (\mu_o - \frac{1}{2}(m+\nu)) \tau} + c_2 e^{-i (\mu_o + \frac{1}{2}(m+\nu)) \tau}, \\
    a_{x+1} &= c_3 e^{-i(\mu_o+\frac{1}{2}(m-\nu)) \tau} + c_4 e^{+i (\mu_o-\frac{1}{2}(m-\nu)) \tau}.
  \end{align}
\end{subequations}
Prefactors $c_i$ depend on parameters $\nu+m$, $JA$, and the initial state of $b_x$ and $a_{x+1}$.  We do not calculate them here explicitly.

\begin{figure*}
    \centering
    \includegraphics[width=\textwidth]{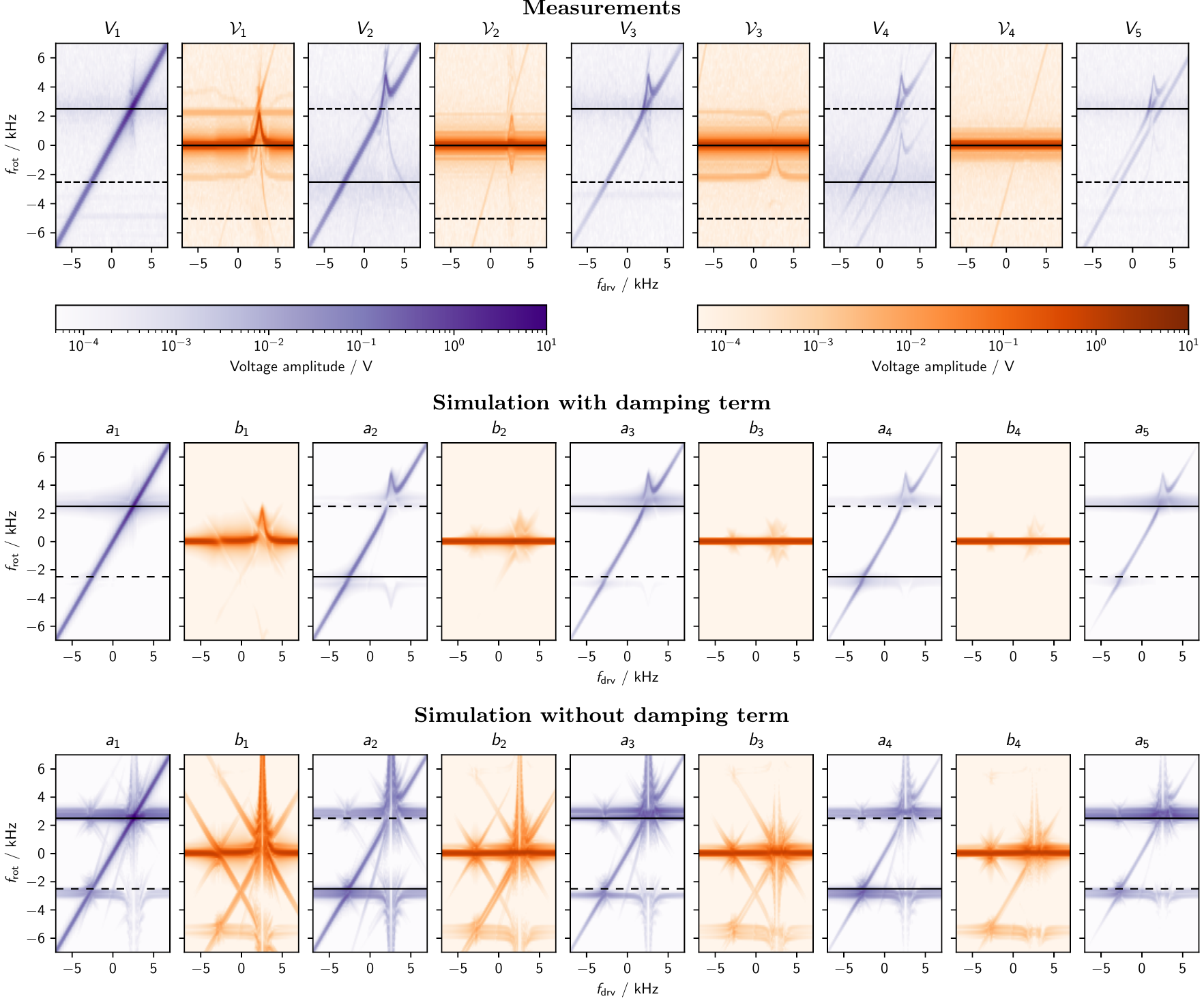}
    \caption{Comparison of measurements and simulations of the driven lattice, which is also the basis for Fig.~\ref{fig:interaction}. As a guide for the eye, the staggered levels from perturbation theory in the large mass limit are indicated in black.  Without driving, amplitudes and the frequency shift diverge when driving on resonance. With the empirical dissipation term, features in the simulation match the measurements.}
    \label{fig:driving-full}
\end{figure*}

The calculation above holds for any given odd $x$ along the chain to calculate the behavior of its following even site.  Next we look at the other case: given a solution $a_x=Ae^{i\nu \tau}$ with even $x$, we calculate the dynamics of the next or previous odd site $a_{x\pm1}$:
\begin{equation}
  \begin{pmatrix} \dot b_x^* \\ \dot a_{x+1} \end{pmatrix}
  = i \begin{pmatrix}
  0 & - J a_x^*(\tau) \\
  J a_x(\tau) & m
  \end{pmatrix}
  \begin{pmatrix} b_x^* \\ a_{x+1} \end{pmatrix}.
\end{equation}
The appropriate variable change is
\begin{subequations}
\begin{align}
    b_x^* &\rightarrow b_x^{*\prime} & b^* &= e^{-i\frac{1}{2}(m-\nu)\tau} b^{*\prime} \\
    a_{x+1} &\rightarrow a_{x+1}^\prime &\qquad a_{x+1} &= e^{+i\frac{1}{2}(m+\nu) \tau} a_{x+1}^\prime
\end{align}
\end{subequations}
to reach the expression
\begin{equation}
  \begin{pmatrix} \dot b_x^{*\prime} \\ \dot a_{x+1}^\prime \end{pmatrix}
  = i \begin{pmatrix}
  -\tfrac{1}{2} (m-\nu) & - J A^* \\
  J A & +\tfrac{1}{2} (m-\nu)
  \end{pmatrix}
  \begin{pmatrix} b_x^{*\prime} \\ a_{x+1}^\prime \end{pmatrix}.
\end{equation}
Eigenvalues of the matrix are
\begin{equation}
    \mu_e(\nu, A) = \pm\frac{1}{2}\sqrt{(m-\nu)^2 - 4 J^2 |A|^2}.
\end{equation}
and the solutions have the frequency components
\begin{subequations}
\begin{align}
    b_x = e^{i(\mu_e + \frac{1}{2} (m-\nu))\tau)} + e^{i(-\mu_e + \frac{1}{2} (m-\nu))\tau)}, \\
    a_{x+1} = e^{i(\mu_e + \frac{1}{2} (m+\nu))\tau)} + e^{i(\mu_e - \frac{1}{2} (m+\nu))\tau)}.
\end{align}
\end{subequations}
The prediction for the frequency shift in Fig.~\ref{fig:interaction} is based on measured site amplitudes $A_x$ shown in Fig.~\ref{fig:interaction-hierarchy} (made dimensionless with scale $V_0=\SI{1}{V}$).  The red line in the spectrum of $V_2$ is
\begin{equation}
    f_{U2}(\nu)/f_0 = \mu_o(\nu, A_1) - (m - \nu)/2 - \nu.
\end{equation}
Predictions for the long range interactions in Fig.~\ref{fig:interaction} use the iterated approach:  The frequency of the third site $\nu_3$ is based on the solution of the second site:
\begin{subequations}
\begin{align}
    \nu_2 &= \mu_o(\nu, A_1) - (m - \nu)/2, \\
    \nu_3 &= \mu_e(\nu_2, A_2) - (m - \nu_2)/2, \\
    \nu_4 &= \mu_o(\nu_3, A_3) - (m - \nu_3)/2, \\
    \nu_5 &= \mu_e(\nu_2, A_2) - (m - \nu_2)/2.
\end{align}
\end{subequations}
The frequency shifts are calculated relative to the driving: $f_U=\nu f_0-\fdrv$.  Fig.~\ref{fig:driving-full} shows the full spectrum from which Fig~\ref{fig:interaction} is calculated.  Also shown are comparisons to numerical simulations of Hamilton's equations in the rotating frame with and without the empirical dissipation term.

\end{document}